# Wavelets and Quantum Algebras

A. Ludu[†], M. Greiner[‡] and J. P. Draayer[†]

[†]*Department of Physics and Astronomy, Louisiana State University,*
*Baton Rouge, LA 70803-4001, U.S.A.*
[‡]*Institut für Theoretische Physik, Technische Universität,*
*D-01062 Dresden, Germany*

**Abstract:**

Wavelets, known to be useful in non-linear multi-scale processes and in multi-resolution analysis, are shown to have a q-deformed algebraic structure. The translation and dilation operators of the theory associate with any scaling equation a non-linear, two parameter algebra. This structure can be mapped onto the quantum group $su_q(2)$ in one limit, and approaches a Fourier series generating algebra, in another limit. A duality between any scaling function and its corresponding non-linear algebra is obtained. Examples for the Haar and B-wavelets are worked out in detail.

# 1  Introduction

Interest in wavelet theory and its applications in multi-resolution analysis [1] has grown over the last decade, involving new and linking disparate fields of research from pure mathematics and physics to down-to-earth signal engineering. It is being widely applied in signal processing and data compression [2-4], pattern recognition [5], statistical physics and turbulence [6], jet dynamics [7], field theory [8,9], solid state physics [10-12], quantum mechanical applications [13-16], non-linear dynamics [6,7,17], soliton theory [18,19], etc.

The rise in interest in wavelet theory is motivated by the fact that Fourier analysis is ineffective when dealing with non-linear models and localize sharp features [6-18]. A central property of wavelets is their ability to expand and analyze functions with respect to a set of self-similar localized basis functions (scaling functions and wavelets) and to processes them locally without affecting the scale. The wavelet method is recursive and therefore ideal for computational applications. Moreover, the scaling functions and the corresponding wavelets are very well localized both in the time and frequency domains, and hence wavelets are ideal for an analysis of phenomena where different space/time scales occur. They also provide mathematical representations that can handle both analytical and numerical difficulties due to singular phenomena [6,7]. Like Fourier analysis, wavelet theory uses basis functions with different characteristic scales. However, Fourier analysis has the advantage of being build upon a simple and solid mathematical foundation [20]. The challenge is to create such a platform for wavelet theory.

This paper is a first step in that direction. It focuses on obtaining a closed algebraic method for the construction of the scaling and wavelet functions. Results for the Haar and B-wavelets are given. The results show that in addition to providing a multi-resolution basis of self-similar functions, wavelets display a definite non-linear symmetry which can be associated with a q-deformation of the Fourier series generating algebra. In a second paper we will show that wavelets also have variational properties: all multi-scale equations follow from Hamilton's equation of an infinite-dimensional Hamiltonian system.

The two-scale equation, which is the central object of wavelet theory, is also shown to have an algebraic structure that is associated with a quantum algebra [21]. Applications of this result can be realized using supercomputers and efficient numerical schemes which, in turn, can be used to model complex continuous systems or genuine discrete systems defined on lattices. The symmetries related to complete solvability [22] or to discrete-continuous transitions [23], and their basic tools, the finite difference equations, are new and key elements in any analysis of discrete systems. On uniform lattices the symmetry algebras of partial differential equations is left unchanged by the discretization [24]. In the case of non-uniform lattices one needs to introduce generalized symmetries, like quantum algebras [25]. This is an example of when the q-deformation of some initial symmetries can play an important role. An outcome of the present paper is an extension of traditional Fourier analysis towards wavelet expansions by means of q-deformation.

The finite-difference and scaling operators involved in the dilation equation are algebraically closed with respect to certain non-linear commutation relations. This symmetry is seen most simply when the operators are realized in terms of q-deformed derivatives [26]. By expressing the translation and dilation operators as q-deformed derivatives, the scaling and difference operators can be mapped onto the generators of a non-linear algebra and the action of the q-deformed derivatives is extended to non-differentiable functions. Moreover, such non-linear algebras can be un-deformed to the Fourier series generating algebra or can be mapped onto the $su_q(2)$ algebra.

After introducing notation and basic definitions in Section 2, we present in Section 3 the deformation of a Fourier algebra into a scaling function generating algebra, and identify a spectral problem with the Haar dilation equation. In Section 4, we introduce the definition of the most general scaling function and wavelet algebra and prove that the algebraic formulation of the dilation equation is unique. A duality relation is obtained: for any scaling function there is a non-linear algebra; and for special algebras one can find a corresponding wavelet structure. In Section 5 we give examples of scaling function algebras for the Haar and the B-wavetes. We use new limiting procedures to solve the q-difference and finite-difference equations. In Section 6 we use a variant and extension of the deforming functional technique [28,29,31] to obtain a mapping of the generators of non-linear algebras to those of $su_q(2)$. Concluding comments, further extensions, and other remarks are given in Section 7.

## 2 Basic elements

An algebraic structure can be built for any scaling function system and wavelet basis. In order to realize this construction, three basic building blocks are needed: the algebraic structure of the Fourier system (starting point), the wavelet theory (final objective), and q-deformation (intermediate tool).

### 2.1 Fourier algebras

In order to demonstrate our approach, consider algebras which have Fourier series as the basis of their representation spaces [20]. In the following we shall denote $\partial^k f/\partial x^k = \partial_k f = f^k$. The trigonometric (Fourier) system $\{|k> = e^{ikx}\}_{k \in Z}$ diagonalizes all translation invariant operators acting on $L^2([0, 2\pi])$. We introduce three generators within a differential realization $J_0 = -i\partial$, $J_\pm = e^{\pm ix}(-i\partial)^p$, satisfying the commutation relations

$$[J_0, J_\pm] = \pm J_\pm, \quad [J_+, J_-] = 1 - (1 - 2i\partial)^p. \tag{1}$$

For $p = 0$, eqs.(1) describe an algebra, denoted $\mathcal{F}_0$, that is isomorphic to an analytical prolongation, $e(2, C)$, of the Euclidean algebra $e(2, R)$ of rigid motions in the plane. The algebra $e(2, R)$, generated by $P_x = P_x^\dagger$, $P_y = P_y^\dagger$ and $R = -R^\dagger$, with the commutators $[R, P_{x,y}] = \pm P_{y,x}$, $[P_x, P_y] = 0$, is realized through the mapping

$$J_0 = iR, \quad J_\pm = P_x \pm iP_y,$$

onto the $\mathcal{F}_0$ algebra. The unitary irreducible representations (unirreps) of $\mathcal{F}_0$ are based on the eigenvectors of the self-adjoint operator $J_0$, $J_0|n> = n|n>$. For any two distinct eigenstates, $|n>$ and $|n'>$, by using the first commutator in eq.(1), we obtain $n' = n + 1$ and hence the spectrum of $J_0$ is unbounded, discrete and consists in equidistant eigenvalues. Hence the space of representations is generated by the Fourier system, and the other generators $J_\pm|n> = |n \pm 1>$ act like ladder operators on the $|n>$ states, increasing/decreasing the scale by unity. For $p = 1$ eqs.(1) describe another Lie algebra, $\mathcal{F}_1$, isomorphic with the symplectic algebra $sp(2, R) \simeq su(1, 1) \simeq so(2, 1)$. This differential realization has the same representation space as $\mathcal{F}_0$ but it is not irreducible. The action of the generators of $\mathcal{F}_1$ is similar to that of the generators of $\mathcal{F}_0$ for $n \neq 0$. Unlike the $\mathcal{F}_0$ case, however, we have that $J_\pm|\mp 1>=|0>$ and the sub-spaces $\{e^{inx}\}_{n\in N}$ and $\{e^{-inx}\}_{n\in N}$ are invariant subspaces of $\mathcal{F}_1$. The $\mathcal{F}_{0,1}$ examples suggest the possibility of other constructions in terms of the operator $\partial$ and the complex exponential functions of different scales.

## 2.2 Scaling functions and wavelets

Wavelets are able to reconstruct a signal through regular sampling, namely, by analyzing the signal at different scales (which increase/decrease exponentially) with the step size between each scale being the same. Different from the ordinary Fourier transform, which reproduces a function as a superposition of complex exponentials, or from the windowed Fourier transform, which introduces a scale into the analysis of signals, multi-resolution analysis processes the signal locally, using the appropriate local scale. Basically, wavelets are constructed with a pair of operators: the dilation (scaling) and finite-difference (combinations of translations) operators, acting on $L^2(R)$ and defined, respectively, by $T^\alpha f(x) = f(x+\alpha)$, $D^\beta f(x) = 2^\beta f(2^\beta x)$, with $\alpha, \beta$ arbitrary real numbers. They are invertible, unitary, and fulfill $T^\alpha D^\beta = D^\beta T^{2^\beta \alpha}$, $T^\alpha T^\beta = T^{\alpha+\beta}$, $D^\alpha D^\beta = D^{\alpha+\beta}$. The formal Taylor series representation of these operators are $T^\alpha = e^{\alpha \partial}$ and $D^\beta = 2^\beta e^{\beta ln 2 x \partial}$. On the space of compact supported or rapidly decreasing functions, any holomorphic function $f(T)$ is locally polynomial. Indeed, if $f$ is holomorphic, and its action is taken on the compact supported function subspace of $\Phi \in L^2(R)$, then the action of $f(T) = \sum_{k \in Z} C_k T^k$ reduces to that of a Laurent polynomial by keeping only a finite number of terms in the sum.

The wavelet system is given by a set of scaled and translated copies of a pair of functions: the scaling function $\Phi$ and the mother wavelet $\Psi$. The basic fact about wavelets is that both these fundamental functions are finite linear combinations of $\Phi$, reflecting the self-similar character of the wavelet system. The defining equation for the scaling function (the dilation equation) is a linear finite-difference equation, including a scale change (q-

difference [27])

$$\Phi(x) = D \sum_{k=0}^{n} C_k T^{-k} \Phi(x) = Dg(T)\Phi(x), \qquad (2)$$

where the RHS sum is a polynomial in $T$, $g(T)$. Eq.(2) is a fixed-point equation and consequently it has only one unique solution [1-7,11,13]. The scaling function $\Phi(x)$, as a solution of eq.(2), is required to have two properties [1-5]:

1. $\int_R \Phi(x)dx = 1$, the average value property;
2. $<\Phi_n, \Phi> \equiv \int_R \Phi(x+n)\Phi(x)dx = \delta_{n,0}$ for any $n \in Z$, the orthogonality condition.

These conditions introduce restrictions on the coefficients $C_k$ in eq.(2) [1-4],

$$\sum_{k=0}^{n} C_k = 2, \quad \sum_{k=0}^{n} C_k C_{k+2l} = \delta_{0l}, \quad l \in Z. \qquad (3)$$

The conditions expressed through eq.(3) yield a pattern of $L^2(R)$ as a chain of subspaces $V_j \subset V_{j+1}$, each one being generated by all the translations of $D^j\Phi$, $j \in Z$. By repeated application of $D$ and $T$ on $\Phi$, $D^j T^n \Phi(x) = \Phi(2^j x + n) \equiv \Phi_{j,n}$, one obtains a non-orthogonal basis in $L^2(R)$. The action of the operator $-DT^{-\lambda}g(-T^{-1})$ on $\Phi$, $\lambda$ being a unique odd integer, provides the mother wavelet function:

$$\Psi(x) = -T^{-\lambda}Dg(-T^{-1})\Phi(x) = -T^{-\lambda}\sum_k (-1)^k C_{-k+1}\Phi(2x-k). \qquad (4)$$

The wavelet $\Psi(x)$ has the property that $\{\Psi_{j,n}\}_{j,n\in Z}$ is an orthonormal basis of $L^2(R)$ [1-3] and $L^2(R)$ is a direct sum of the orthogonal subspaces $W_j$ (the orthogonal complements of $V_j$), each of them generated by all possible translations of $D^j\Psi = \Psi_{j,0}$ with integral m: $L^2(R) = \oplus_{j\in Z} W_j$, [1-5]. Hence translated and dilated copies of the mother wavelet, $\Psi(2^j x + k)$, generate a true orthonormal basis. Recursively applications of the eqs.(2,4) relate all the scaling functions and wavelets.

The simplest example is provided by the Haar wavelet, defined by the scaling function $\Phi_{Haar}(x) = 1$ if $|x - 1/2| \leq 1/2$ and 0 otherwise. In this case we have $D(1 + T^{-1})\Phi_{Haar} = \Phi_{Haar}$ and $g_{Haar}(T) = 1 + T^{-1}$ ($C_0 = C_{-1} = 1$ and the rest 0 in eq.(2)). The corresponding Haar wavelet is $\Psi_{Haar} = D(1 - T^{-1})\Phi_{Haar}$, i.e. $\lambda = 1$ in eq.(4).

There are similarities and differences between the Fourier and the wavelet approaches. From an algebraic point of view, in both approaches, there are eigenfunction equations, which keep the scale constant, and the ladder operators, which change the scale. The dilation equation, eq.(2), has its analog in the Fourier formalism, though it is a fixed-point equation and has only one solution. The mother wavelet, eq.(4), has no analog in Fourier analysis. Each Fourier eigenfunction carries one scale; a wavelet function $\Psi(x)$ involves two scales, e.g. $\Phi_{0,k}$ and $\Phi_{2,k}$. Because of their localization, wavelets possess a degree of freedom beyond that of a Fourier system; namely, the width of the support of $\Phi$. In the present approach this degree of freedom is associated with the deformation parameter, $q$.

## 2.3 Quantum deformations

A possibility for constructing scaling functions/wavelets within an algebraic approach is to deform the Fourier algebraic structure into a non-linear system. In general, the scaling functions and wavelets are not differentiable. This suggests the use of finite-difference operators instead of derivatives. Finite-difference operators are closed with respect to commutation but only within non-linear algebraic constructions. Consequently, it is natural to use q-deformed derivatives to find a foundation for wavelet theory in $q$-deformed algebras.

Quantum algebras refer to some specific deformations of Lie algebras, to which they reduce when the deformation parameter $q$ is set equal to unity (for a recent monograph see [21] and references therein). The simplest example of a $q$-algebra is $su_q(2)$ whose Jordan-Schwinger realization is given in terms of $q$-bosonic operators [26]. The q-deformed algebras have been applied in various branches of physics like spin-chain models, non-commutative spaces, rotational spectra of deformed nuclei, Hamiltonian quantization, and dynamical symmetry breaking. The basic element is the q-deformation of a certain object $x$, which can be a number, an operator, or a function:

$$[x]_s = \frac{q^x - q^{-x}}{q - q^{-1}} \xrightarrow{s \to 0} x, \qquad (5)$$

where $q = e^s$. The Taylor expansion of $T$ and $D$ are related to the q-deformation of the derivative operator, according to the definition of the coordinate description of the q-deformed oscillator introduced in [26],

$$[\partial]_s f = \frac{T^s - T^{-s}}{2 sinh(s)} f(x). \qquad (6)$$

Analogously, we introduce the operator

$$[x\partial]_{s\ln 2}f(x) = \frac{f(2^s x) - f(2^{-s} x)}{2sinh(s\ln 2)} = \frac{D^s - D^{-s}}{2sinh(s\ln 2)}f(x). \qquad (7)$$

When $s \to 0$, $[\partial]_s \to \partial$ and $[x\partial]_{s\ln 2} \to x\partial$. Eqs.(6,7) can be inverted and the occurrence of the q-deformed derivative and the translation/dilation operators becomes immediate

$$T^{\pm s} = \frac{1}{2}\left(\pm \frac{[\partial]_s}{\eta(s)} + \frac{[\partial]_{s/2}^2}{\eta^2(s/2)} + 2\right) \qquad (8)$$

$$D^{\pm s} = \frac{1}{2}\left(\pm \frac{[x\partial]_s}{\eta(s\ln 2)} + \frac{[x\partial]_{s/2}^2}{\eta^2(s\ln 2/2)} + 2\right), \qquad (9)$$

where $s \in N$ (or more general $\in R$) and $\eta(s) = \frac{1}{e^s - e^{-s}}$. The q-deformed algebra $su_q(2)$, generated by $\{J_0, J_\pm\}$, is defined by the q-deformed version of the commutation relation

$$[J_+, J_-] = [2J_0]_s$$

while the other two commutator relations remain undeformed, that is, they have the same form as for $su(2)$. In addition to this traditional version of $su_q(2)$, several generalized forms have been introduced through two different prescriptions: 1) by deforming the commutator $[J_0, J_\pm]$ by using some arbitrary function $G(J_0, q)$, $[J_0, J_+] = G(J_0, q)J_+$ and $[J_0, J_-] = -J_- G(J_0, q)$, independently proposed in [28], [29] and [30]; and 2) by deformations involving all three commutation relations, using two functions $G(J_0, q)$ and $F(J_0, q) = [J_+, J_-]$, introduced in [31]. Unlike the former, for which the spectrum of $J_0$ is linear, the latter is characterized by an exponential spectrum for $J_0$. Since in wavelet theory the domains of analysis are divided exponentially, rather than linearly with bands of equal widths, such algebras are used in the following analyses.

## 3  Haar scaling function algebra, $\mathcal{A}_{s,\alpha}$

The aim in this section is to obtain operators depending on $D$ and $T$ that form a non-linear algebra. This structure must provide an algebraic form for eqs.(2,4) and map onto the Fourier algebra, $\mathcal{F}_{0,1}$. Consider an operator depending on $T$ and two real parameters $s, \alpha$:

$$W_0(s, \alpha) = \frac{T^{s\alpha} - T^{-s\alpha} \cos s\pi}{2\xi(\alpha) \sinh s}, \tag{10}$$

where $\xi(\alpha) = \frac{1}{\sinh(1)} \sin(\frac{\alpha\pi}{2}) - 2i \cos(\frac{\alpha\pi}{2})$. The $(s, \alpha)$ parameters allow $W_0$ to approach $\partial$ or combinations of $T$

$$W_0(0, \alpha) = \frac{\alpha}{\xi(\alpha)} \partial, \tag{11}$$

$$W_0\left(\frac{1}{2}, \alpha\right) = \frac{1}{2\sinh(1/2)\xi(\alpha)} T^{\frac{\alpha}{2}}, \tag{12}$$

$$W_0(1, \alpha) = \frac{T^\alpha + T^{-\alpha}}{2\sinh(1)\xi(\alpha)}, \tag{13}$$

$$W_0(2, \alpha) = \frac{T^{2\alpha} - T^{-2\alpha}}{2\sinh(2)\xi(\alpha)}. \tag{14}$$

In the limit $s \to 0$, $W_0$ reduces to the normal derivative with respect to $x$, eq.(11). In the limit $s = \frac{1}{2}$, we obtain a power of the translation operator $T$, eq.(12). Eq.(13) defines a q-deformation of unity, namely, $W_0(1, 0) = \frac{i}{4\sinh(1)}$. $W_0(s, \alpha)$ can be a Hermitian or anti-Hermitian operator, $W_0(1, \alpha) = \pm W_0(1, \alpha)^\dagger$, depending on $\alpha$. The last limit, eq.(14), represents a finite-difference operator and is proportional to the q-derivative with respect to the $\alpha$-deformation $[2\partial]_\alpha$, similar to eqs.(6-9). For $\alpha = 2k$ in eq.(12), $\alpha = k$ in eq.(13) and $\alpha = k/2$ in eq.(14), the corresponding $W_0(s, \alpha)$ are Laurent polynomials in $T$, providing a direct connection to the wavelet operators. For some values of $\alpha$, eqs.(12-14) give invertible operators with respect to $T$, $T = T(W_0)$.

In order to introduce the dilation operator $D$, we define

$$W_\pm(s) = \frac{1}{2} e^{\mp ix\frac{s-1}{2}} D^{\mp s} e^{\mp ix\frac{s-1}{2}} (1 + T^{\pm s}). \tag{15}$$

In order to fulfill the hermiticity condition $(W_+)^\dagger = W_-$ on $L^2(R)$ these generators can be redefined in the form

$$W_\pm \to \tilde{W}_\pm = 2^{-1 \mp \frac{s}{2}} e^{\mp ix\frac{s-1}{2}} D^{\mp s} e^{\mp ix\frac{s-1}{2}} \left(1 + e^{-i\frac{s(s-1)(1+2^s)}{4}} T^{\pm s(2^{\frac{s}{2}(1\mp 1)})}\right).$$

By using the commutators relations between $D$ and $T$ and by performing an integration by parts, one has a direct check of the relation $(\tilde{W}_+)^\dagger = \tilde{W}_-$

$$< f_1, \tilde{W}_+ f_2 > = \int_R f_1^*(x) \tilde{W}_+ f_2(x) dx = < \tilde{W}_- f_1, f_2 > .$$

From eqs.(10,15) one obtains the commutators

$$[W_0(s,\alpha), W_+(s)] = G(s,T)W_+(s), \qquad (16)$$

$$[W_0(s,\alpha), W_-(s)] = -W_-(s)G(s,T), \qquad (17)$$

$$[W_+(s), W_-(s)] = F(s,T), \qquad (18)$$

with

$$G(s,T) = W_0(s,\alpha) - \frac{T^{2^s s\alpha} e^{is\alpha(1+2^s)\frac{s-1}{2}} - T^{-2^s s\alpha} e^{-is\alpha(1+2^s)\frac{s-1}{2}} \cos s\pi}{2\sinh(s)\xi(\alpha)} \qquad (19)$$

and

$$F(s,T) = \frac{1}{4}\Big((1 + e^{is(1+2^s)\frac{s-1}{2}} T^{2^s s})(1+T^{-s}) - (1 + e^{is(1+2^{-s})\frac{s-1}{2}} T^{-2^{-s}s})(1+T^s)\Big). \qquad (20)$$

Eqs.(10,15,16-20) describe a non-linear associative algebra, generated by $T^\pm, D^\pm$, denoted $\mathcal{A}_{s,\alpha}$. The eigenvalue problem for $W_\pm$ provides the algebraic form for the scaling equation. When $W_0(s,\alpha)$ is invertible with respect to $T$, the algebra has $W_0, W_\pm$ as generators which depend on two parameters $(s,\alpha)$ and is homomorphic with a special q-deformed algebra, namely, the two-color quasitriangular Hopf algebra $\mathcal{A}^\pm$ [31]. The spectrum of $W_0$ is not equidistant because $F(s,T) \neq const. \times W_0$. The Casimir operator of the algebra $\mathcal{A}_{s,\alpha}$ is given by [28-31]

$$C = W_-W_+ + H(W_0) = W_+W_- + H(W_0) - F(W_0), \qquad (21)$$

where $H$ is a real function, holomorphic in a neighborhood of 0 and which must satisfy the functional equation

$$H(\xi) - H(\xi - G(\xi)) = F(\xi). \qquad (22)$$

with $\xi$ generic. The irreps of $\mathcal{A}_{s,\alpha}$ are labeled by the eigenvalues $W_0|a> = a|a>$. The commutator in eq.(16) can be written in the form

$$W_+W_0 = (W_0 - G)W_+. \qquad (23)$$

If $|a> \neq |a'>$ are eigenvectors of $W_0$, with corresponding eigenvalues $a, a'$, we have from eq.(23) a non-linear recursion relation for all the eigenvalues

$$a' = a - G(a). \qquad (24)$$

The algebra $\mathcal{A}_{s,\alpha}$ can be mapped into $su(1,1)$, the Lie algebra of a Fourier series, in the limit $s \to 0$, $\alpha \to 2$: $W_\pm \to e^{\pm ix} = J_\pm$, $W_0 \to -i\partial = J_0$, where we take these limits in the order $\mathcal{A}_{s_0,\alpha_0} = \lim_{\alpha \to \alpha_0} \lim_{s \to s_0} \mathcal{A}_{s,\alpha}$. The continuous mapping $\mathcal{A}_{s,\alpha} \to \mathcal{A}_{0,2} \simeq su(1,1)$ is an algebraic morphism and one can consider wavelets as q-deformed generalizations of the Fourier series, in the above sense.

As a first example, the Haar scaling function can be associated with a particular case of $\mathcal{A}_{s,\alpha}$, namely, $\mathcal{A}_{1,1}$. In the algebra $\mathcal{A}_{1,1}$ the eigenproblem for $W_-$ provides the dilation equation for the Haar scaling function $\Phi(x)$ defined by eq.(2) with $g(T) = 1 + T^{-1}$. We have in this case $G = W_0 - 2W_0^2 + 1$, and $W_\pm = \frac{1}{2}D^{\mp 1}(1 + T^{\pm 1})$ and the commutators

$$[W_0, W_+] = (W_0 - 2W_0^2 + 1)W_+,$$

$$[W_0, W_-] = -W_-(W_0 - W_0^2 + 1), \quad (25)$$

$$[W_+, W_-] = \frac{1}{4}\Big(T^2(W_0) - T^{1/2}(W_0) - T^{-1/2}(W_0) + T^{-1}(W_0)\Big),$$

where $T(W_0)$ is the solution of the equation $2W_0 = T + T^{-1}$, a pseudo-differential operator. The Casimir operator of $\mathcal{A}_{1,1}$ is a constant. Indeed, eq.(22) for the general Casimir operator is given by

$$H(W_0) - H(2W_0^2 - 1) = F(W_0), \quad (26)$$

and has, in terms of the $T = T(W_0)$ operator, the form

$$\chi(T) - \chi(T^2) = \frac{1}{4}\Big(T^2 - T^{1/2} - T^{-1/2} + T^{-1}\Big), \quad (27)$$

where $\chi(T) = H(\frac{T+T^{-1}}{2})$. The unique analytical solution for eq.(27) reads

$$\chi(T) = \frac{1}{4}\Big(const. - T - T^{1/2} - T^{-1/2}\Big), \quad (28)$$

which gives for the Casimir operator a constant.

The spectrum of $W_0$ consists of periodic functions satisfying the equation $\Phi(x+1) + \Phi(x-1) = 2\Phi(x)$. This takes one back to Fourier analysis. The way to the Haar scaling function is to use the spectrum of $W_\pm$ since in $\mathcal{A}_{1,1}$ the eigenproblem for $W_-$ gives the dilation equation for the Haar scaling function $\Phi(x)$. Once $\Phi$ is obtained from $2W_-\Phi = \Phi$ the generator $W_0$ carries $\Phi$ into an infinite sequence of functions $W_0^n \Phi$. These functions are mutually orthogonal and generate the space $V_0$. The action of $W_\pm$ on $\Phi$ yields $W_0^{2^n}\Phi$. Consequently, $V_0$ is an invariant space for all the generators of the algebra $\mathcal{A}_{1,1}$.

The spectrum of $W_0$ depends on the values of the parameters $s, \alpha$. In the case of $\mathcal{A}_{1,1}$ this relation is $a' = 2a^2 - 1$. The corresponding representations are infinite-dimensional and the eigenvalues of $W_0|_{1,1}$ are $a_n = cosh(2^n \tilde{a})$, $n \in Z$ for any $\tilde{a}$. The corresponding eigenfunctions have the form $|a_n>_\pm = e^{\pm 2^n \tilde{a} x}$ which results in an exponential spectrum for $a_n$ similar to the sequence of scales in wavelet theory. This is a self-similar spectrum with respect to $\tilde{a}$, like the sequence of scales in wavelet theory. A part of this spectrum is shown in Fig. 1. The action of $W_\pm$ is:

$$W_\pm |a_n>_\pm = \frac{1 + e^{\pm 2^n \tilde{a}}}{2} |a_{n\pm 1}>_\pm . \tag{29}$$

Another basis for a representation of $\mathcal{A}_{1,1}$ is given by $|k> = e^{\frac{i\pi x}{k}}$, $k \in Z$. We have the action

$$W_0|k> = \cos\frac{\pi}{k}|k>, \quad W_\pm|k> = \frac{1}{2}\left(1 + e^{\pm \frac{i\pi}{k}}\right)|2^{\pm 1}k>,$$

and on this basis $(W_+)^\dagger = W_-$. There are two invariant spaces, $|2^k>$ and $|2^{-k}>$, $k \in N$, with $W_\pm |0> = 1$.

The same procedure can be followed for any set of the parameters $s, \alpha$. For example, if we choose the algebra $\mathcal{A}_{2,1/2}$ with $W_0$ defined in eq.(12), we obtain the spectrum of $W_0$ described by the recursion relation:

$$a' = a\left(a + \sqrt{a^2 + 1} + \frac{1}{a + \sqrt{a^2 + 1}}\right), \tag{30}$$

which also gives a non-linear, unbounded, discrete representation of $\mathcal{A}_{2,1/2}$.

The last thing to prove is the uniqueness of the two-scale equation in the $\mathcal{A}_{1,1}$ scaling function generating algebra. The two-scale equation, eq.(2), in its algebraic form $2W_-\Phi = \Phi$ is not unique in $\mathcal{A}_{1,1}$ if there exists an operator, similar with that occuring in the dilation equation, which commutes with $W_-$. This operator should contain higher powers in $T$ and consequently is an element of $U(\mathcal{A}_{1,1})$.

**Proposition 1**: *In $U(\mathcal{A}_{1,1})$ there exists a unique operator $X = Dx(T^{-1})$ such that $[W_-, X] = 0$. The function $x(\xi)$ is integer, not polynomial, and unbounded, for generic $\xi$.*

**Proof:** We take for $X$ a Laurent series $x(T^{-1}) = \sum_{k \in Z} C_k T^{-k}$ then the condition $[W_-, X] = 0$ results in a recursion relation for the coefficients $C_k$:

$$C_{2k+1} = C_k - C_{2k-1}, \quad C_{2k} = C_k - C_{2k-2}, \tag{31}$$

for any $k \in Z$. Eqs.(31) have one trivial solution which reproduces $W_-$, $x(T^{-1}) = C_{-1}(1 + T^{-1})$ and only one other solution, with $C_k = \pm C_{-1}$ or 0, uniquely defined ($C_2 = C_{\pm 3} = C_{-4} = C_5 = C_{-6} = C_7 = C_{-8} = C_{-11}... = 0$, $C_{-2} = C_{-5} = C_6 = C_{-9} = C_{-10} = C_{10} = C_{11} = ... = C_{-1}$ and $C_1 = C_4 = C_{-7} = C_8 = C_9 = C_{-12} = C_{12} = ... - C_{-1}$, etc.). The sequence of non-zero coefficients is infinite, and $\{C_k\}$ is not a Cauchy sequence. (q.e.d.) We note that the $\mathcal{A}_{1,1}$ algebra is also a Hopf algebra, [21,30,31], defined by eqs.(25) and by the coproduct, counit and antipode in the form:

$$\triangle T^{\pm} = T^{\pm} \otimes T^{\pm}, \quad \triangle D^{\pm} = D^{\pm} \otimes D^{\pm}, \quad \epsilon(T^{\pm}) = \epsilon(D^{\pm}) = 1,$$

$$S(T^{\pm}) = T^{\mp}, \quad S(D^{\pm}) = D^{\mp}.$$

All its generators are primitive elements.

## 4  General scaling and wavelet algebra

This algebraic approach for the Haar scaling function can be generalized to yield a non-linear algebra for any scaling function, and conversely, to find the dilation equation and scaling function for certain algebras. The procedure starts with an algebra generated by the dilation and translation operators and constructs, within this algebra, the two-scale equation. The generators of the $\mathcal{A}_{s,\alpha}$ algebra can be generalized as

$$W_0 \to j_0(T), \quad W_{\pm} \to j_{\pm}(D,T) = e^{\mp ix\frac{s-1}{2}} D^{\mp s} e^{\mp ix\frac{s-1}{2}} j(T^{\pm 1}), \tag{32}$$

with $j_0(T)$ and $j(T)$ being arbitrary functions of $T$ and $s$, holomorphic in a neighborghood of $T = 1$, with their dependence on $s$ being such that in the limit $s \to 0$, $j_0(T) \to -i\partial$, $j(T) \to 1$. In this case the commutation relations, eqs.(16-20), become

$$[j_0, j_+] = \tilde{G} j_+, \quad [j_0, j_-] = -j_- \tilde{G}, \quad [j_+, j_-] = \tilde{F} \tag{33}$$

with $\tilde{G}, \tilde{F}$ depending on $T$ through $j_0, j$, respectively,

$$\tilde{G} = \tilde{G}(s,T) = j_0(s,T) - j_0\left(s, T^{2^s} e^{i(1+2^s)\frac{s-1}{2}}\right), \tag{34}$$

$$\tilde{F} = \tilde{F}(s,T) = j(s, e^{i\frac{s-1}{2}(1+2^s)}T^{2^s})j(s,T^{-1}) - j(s,T^{-2^{-s}}e^{i\frac{s-1}{2}(1+2^{-s})})j(s,T). \qquad (35)$$

Eqs.(33-35) define a non-linear algebra denoted $\mathcal{A}_{j_0,j}$ as a generalization of $\mathcal{A}_{s,\alpha}$. If the function $j(T)$ is invertible with respect to $T$, $\mathcal{A}_{j_0,j}$ can be expressed in terms of the generators $j_0, j_\pm$ only and then $\tilde{G}(T) = \mathcal{G}(j_0)$, $\tilde{F}(T) = \mathcal{F}(j_0)$. The algebraic morphism $j_0(T) \to W_0$ and $j(T) \to 1 + T^s$ provides the mapping $\mathcal{A}_{j_0,j} \to \mathcal{A}_{s,\alpha}$. Since the first two commutators of $\mathcal{A}_{j_0,j}$, eq.(33), do not depend on the function $j(T)$, and the third commutator in eq.(33) does not depend on the function $j_0(T)$, the closure condition for $\mathcal{A}_{j_0,j}$ is independent of the functions $j_0(T)$ and $j(T)$ and hence they can be choosen in a convenient way to provide any dilation equation in the form of the eigenproblem for $j_-$. The algebraic closure conditions for $\mathcal{A}_{j_0,j}$, together with the definitions of $j, j_0$, require that

$$j_0(T^2) - j_0(T) = \mathcal{G}(j_0(T^2)), \qquad (36)$$

$$j(T^2)j(T^{-1}) - j(T^{-1/2})j(T) = \mathcal{F}(j_0(T)). \qquad (37)$$

Both eqs.(36,37) are non-linear, and in general, difficult to solve analytically. The unique solution for the dilation equation are $j_\pm$ since the Casimir operator of this algebra depends on $T$ only. Choosing $g(T) = j(T)$ yields the dilation equation in the form $2j_-\Phi = \Phi$. In the limit of $\mathcal{A}_{s,\alpha}$ this provides again the Haar two-scale equation. The corresponding scaling function belongs to the basis of the representation of $j_-$ with eigenvalue $1/2$. The remaining arbitrary function $j_0$ can now be selected to obtain the wavelet generating operator $j_-j_0\Phi = \Psi$, in agreement with eq.(4). One can formally construct a basis of the representation for $j_-$ with the functions $|n> = (\ln T)^n \Phi$. The procedure is the following: for a given algebra $\mathcal{A}_{j_0j}$ (therefore given functions $\mathcal{G}, \mathcal{F}$) solve eqs.(36,37) with respect to the functions $j_0, j_\pm$ and obtain the corresponding dilation equation in the form $2j_-\Phi = \Phi$. A certain combination of generators produces the corresponding wavelet equation for $\Psi$. Conversely, given a scaling/wavelet function, and consequently its dilation equation and $g(T)$, one can choose $j_- = Dg(T)$ and then solves the equation $Dg(-T^{-1}) = j_-j_0$ with respect to $j_0$. With $j_0, j_\pm$ known one can solve eqs.(36,37) with respect to $\mathcal{G}, \mathcal{F}$ to construct the algebra.

This algebraic approach is in some sense universal. What is modified is the specific realization of the non-linear algebra in terms of $D$ and $T$. Constraints are imposed by the two limiting approaches (two-scale equation and Fourier limit). Loosley speaking, the closure of the algebra provides the fixed-point dilation equation and the non-linearity of the algebra provides the exponential scaling.

We also note that $V_0 = \{\sum_k C_k T^k \Phi\}$, that is, the space generated by all the integer translations of $\Phi$, and any other $V_j = D^j V_0$, is not invariant to the action of $j_-$. For example, the condition $j_- D\Phi = \sum_k C_k T^k D\Phi$ implies the existence of an operator containing $D$, whose commutator with $j_-$ is a function of $T$ only. From the definition of the algebra this is impossible, and consequently this proves the above conjecture. The $V_j$ spaces form, by recursion, a basis in $L^2(R)$.

In order to prove the uniqueness of the two-scale equation in the general case, we again use Proposition 1, for $W_- \to j_- = Dj(T^{-1}) = D\sum_{k\in Z} j_k T^k$. We have to solve the commutation equation $[Dj(T^{-1}), X] = 0$ for $x(T) = \sum_{k\in Z} X_k T^k$, $X = Dx(T)$. This implies that the arbitrary functions $x(\xi)$ and $j(\xi)$ must fulfill the functional condition ($\xi$ a generic variable)

$$x(\xi^2)j(\xi) = x(\xi)j(\xi^2). \tag{38}$$

Eq.(38) does not carry any restriction with respect to the values of the functions between 0 and 1. Since $Dj(T)$ represents a scaling operator we have, according to the dilation equation, $j(1) = 1$, $j(-1) = 0$ which implies $x(-1) = 0$. If $\Phi$ is the corresponding scaling function for $j_-$ then $X\Phi$ is also an eigenfunction of $j_-$, $j_-(X\Phi) = X\Phi$. In order for $Dx(T)$ to satisfy the average value property (1) in subsection 2.2, we impose the additional condition $x(1) = 1$. The last condition require of for $DX(T)$, the orthogonality condition (2) of subsection 2.2, is equivalent with the condition [1-4]

$$|x(T)|^2 + |x(-T)|^2 = 1. \tag{39}$$

By introducing eq.(38) in eq.(39) we obtain $X = j_-$ which provides the trivial indentity solution, and hence uniqueness.

# 5 Construction of the scaling and wavelet algebra

The link between the scaling function and wavelet, and the corresponding scaling algebra is supported by the solutions of eqs.(36,37). In the following we present a method for solving these equations, from the wavelets towards its algebra. The dilation equation eq.(2) has the form of a fixed-point equation. Therefore we must try to find its eigenvectors as limits of some functional sequences. The limits of these sequences should be compact supported or rapidly decreasing functions for two reasons: to obtain scaling functions with good localization, and to provide correct behaviour of the action of the operators $f(T)$. Since the scaling function will be expressed as a limit of a sequence, one has to look for operators which commute with the limit (are absorbed in the limit). We choose a test function $\triangle_0(x)$ and a sequence of operators $f_n$ from the universal covering $U(\mathcal{A}_{j_0,j})$ of the scaling function algebra introduced in the preceding section, such that the limit $\Phi(x) = \lim_{n \to \infty} f_n \triangle_0(x)$ exists in the weak topology [22] and it provides the scaling function.

In the following we show how to construct the scaling algebra $\mathcal{A}_{j_0,j}$ from the dilation equation, written in the form $j_-(T, D, s)\Phi = \Phi$. We have the following proposition in the framework of the algebra $\mathcal{A}_{j_0,j}$, for $s = 1$:

**Proposition 2** *Let $f_n(j_0)$ be a functional sequence in $U(\mathcal{A}_{j_0,j})$, $s = 1$, $\triangle_0(x)$ a (test) function and $\Phi(x)$ the scaling function. If the limit $\Phi(x) = \lim_{n \to \infty} f_n(j_0) j_-^n \triangle_0(x)$ exists, and $\lim_{n \to \infty} \frac{f_n(j_0 - \mathcal{G}(s,j_0))}{f_{n-1}(j_0)} = 1$, then $j_-\Phi(x) = \Phi(x)$.*

**Proof:**

From the RHS of the first condition in the hypothesis and from eq.(33) we have

$$f_n(j_0) j_-^n = j_- f_n(j_0 - \mathcal{G}(s, j_0)) j_-^{n-1} = j_- \frac{f_n(j_0 - \mathcal{G}(s, j_0))}{f_{n-1}(j_0)} f_{n-1}(j_0) j_-^{n-1}.$$

It follows:

$$j_-\Phi = \lim_{n \to \infty} j_- f_n(j_0) \triangle_0(x) = \Phi,$$

q.e.d. It follows from Proposition 2 that for a given dilation equation $j_-\Phi = \Phi$, (both $j_-$ and $\Phi$ given) we can find out a sequence $f_n(j_0)$, the operator $\mathcal{G}(s, j_0)$, and a test function $\triangle_0(x)$, such that these objects satisfy the hypothesis of Proposition 2. Then it follows that one can construct the scaling algebra, since with $j_0$ and $\mathcal{G}(s, j_0)$ found, $\mathcal{F}(s, j_0)$ results from the last commutator in eq.(33). In general, one starts with $j_0$ as an arbitrary function of $T$ which can be deformed into $-i\partial$ when mapping $\mathcal{A}_{j_0,j}$ to $\mathcal{F}_{0,1}$. As $j_-$ is provided by the two-scale equation, $j_-\Phi = \Phi$, other solutions are forbidden. If $\lim_{n\to\infty} f_n(j_0) = f_\infty(j_0) \neq const.$, this limit should commute with $j_-$ which is forbidden by Proposition 1. This procedure closes the construction of the algebra $\mathcal{A}_{j_0,j}$.

In the following we give an algorithm for finding $f_n(j_0)$ and $j_0(T)$ and illustrate it with two examples. Since any mother scaling function is defined by the dilation equation ($Dh(T)\Phi = \Phi$ or $j_-\Phi = \Phi$) it is natural to search for solutions by using a recursion algorithm. For instance, we can find a test function $\triangle_0(x)$ and a triplet of operators $A, B, C$ such that the limit $\Phi(x) = B \lim_{n\to\infty} A^n \triangle_0(x)$ exists, and in addition, $CB = BA^k$ for a finite positive integer $k$. Then, we have the property $C\Phi(x) = \Phi(x)$. Indeed, $C\Phi = CB \lim_{n\to\infty} A^n \triangle_0 = B \lim_{n\to\infty} A^{n+k} \triangle_0 = B \lim_{n\to\infty} A^n \triangle_0 = \Phi$. From the dilation equation we know that $C$ should have the form $C = Dc(T)$ with $c(T)$ a function of $T$. One of the simplest choices is to use $A = D$ and $B = b(T)$. Then we have $CB = Dc(T)b(T) = c(T^{1/2})b(T^{1/2})D = b(T)A$; that is, $k = 1$ and a restriction for the arbitrary functions $c(T)$ and $b(T)$ arises:

$$c(T^{1/2})b(T^{1/2}) = b(T) \tag{40}$$

This equation is useful in both directions (algebraic $\leftrightarrow$ dilation equation), since one can start with a given scaling function (given $c(T)$) and find the operator $b(T)$ involved in the algebra and conversely.

Now consider the algorithm for the Haar scaling function, $c(T) = 1 + T^{-1}$. We look for solutions of eq.(40) as Laurent series for $b(T) = \sum_{k\in Z} b_k T^k$. In this case eq.(40) has an unique solution, $b(T) = const. \times (1 - T^{-1})$. This gives again an unique solution for $\Phi = (1 - T^{-1}) \lim_{n\to\infty} D^n \triangle_0$. We stress that $Db(T)$ is exactly the operator which gives the Haar wavelet. Further, we can express $\Phi_{Haar}$ in the form $\Phi(x) = H(x) - H(x-1)$, where $H(x)$ is the Heaviside distribution. For any sequence of $C^\infty$ functions $\delta_n(x) \to \delta(x)$, we have $\triangle_n(x) = \int \delta_n(x)dx \to H(x)$ and we can write

$$\Phi(x) = \lim_{n\to\infty} (\triangle_n(x) - \triangle_n(x-1)). \tag{41}$$

For example, we can use the sequences $\delta_n = \frac{1}{\pi}\frac{2^{-n}}{x^2+2^{-2n}}$ or $\delta_n = \frac{2^{n-1}}{\cosh^2(2^n x)}$ and $\Delta_n = \frac{\arctan(2^n x)}{\pi}$ or $\Delta_n = \frac{1}{2}\tanh(2^n x)$, respectively. The latter example is a soliton-like shape, having good localization. These sequences converge to $\delta(x)$, respectively. In this way we have selected subsequences that step in powers of 2. Eq.(41) can be written as:

$$\Phi(x) = \lim_{n\to\infty}(1 - T^{-1})D^n \Delta_0(x), \tag{42}$$

We can express this definition in terms of the algebra $\mathcal{A}_{1,1}$. By using eq.(25) and the properties of the operators $D$ and $T$, we can write eq.(42) in the form

$$\Phi(x) = \lim_{n\to\infty}(1 - T^{-2^{-n}})(2W_-)^n \Delta_0(x). \tag{43}$$

Indeed, from the commutation relation between $D$ and $T$ and the definition of $W_-$ we have

$$(2W_-)^n = (D(1 + T^{-1}))^n = D^n(1 + T^{-1} + ... + T^{-2^n+1}) \tag{44}$$

$$= (1 + T^{-2^{-n}} + (T^{-2^{-n}})^2 + ... + (T^{-2^{-n}})^{2^n-1})D^n.$$

Hence, we can write $(1 - T^{-1})D^n = (1 - T^{-2^{-n}})(2W_-)^n$. Moreover, we can write, by using the inverted form for $W_-(T)$, the dilation equation in a pure algebraic form:

$$\Phi(x) = \lim_{n\to\infty}(1 - T^{-2^{-n}}(W_0))(2W_-)^n \Delta_0(x). \tag{45}$$

From this last equation it follows that $2W_-\Phi = \Phi$, i.e., the Haar dilation equation. If the function $\Delta_0(x)$ is choosen from a class of suitable functions for a wavelet analysis then its scaling function $\Phi(x)$, eq.(45), gives a rapidly convergent wavelet expansion. In order to demonstrate this procedure, we show in Fig. 2 scaling functions obtained in this way for different values of the parameter $s$ within $\mathcal{A}_{s,\alpha}$. For $s = 0$ and 1 this functions yields the Haar scaling function.

We give another example for the $\Phi_2$ B-scaling function and the $\Psi_2$ B-wavelet. We define

$$\Phi_2(x) = 4x \text{ for } 0 < x < 1/2; \quad -4x + 4 \text{ for } 1/2 < x < 1, \tag{46}$$

and 0 in the rest, or

$$\Phi_2(x) = \lim_{n\to\infty}\frac{1}{2^{-n+1}}(1 - 2T^{-1/2} + T^{-1})D^n \Delta_{0,tri}(x), \tag{47}$$

with

$$\triangle_{0,tri}(x) = \frac{2x}{\pi}arctan(2x) - \frac{\ln(1+4x^2)}{2\pi} = \int_0^{2\pi} \triangle_0(x)dx \qquad (48)$$

Following the above algorithm we obtain the corresponding dilation equation

$$j_-\Big|_{tri} \Phi_2(x) = \frac{D}{2}(1+T^{-1/2})^2\Phi_2(x). \qquad (49)$$

By using the commutation relation $\partial D = 2D\partial$ and the dilation equation eq.(49) we can obtain a self-contained equation for the wavelet:

$$\Psi_{tri}(x) = D(1 + 2T^{-1/2} + T^2)^2 \Psi_{tri}. \qquad (50)$$

The corresponding algebras for Haar and $\Phi_2$ B-scaling functions are different. For the B-wavelet we have, in $\mathcal{A}_{1,1}$

$$G(W_0) = W_0 - W_0^2 - \frac{1}{2} \qquad (51)$$

This new algebra fulfills different commutation relations and consequently has a spectrum for $W_0$ which is different from that for the Haar scaling algebra. The algorithm introduced above can also be used to generate other scaling functions and their corresponding algebras. An interesting track is to obtain polygon-like wavelets or, by following the same procedure which guided us to obtain the B-scaling function from the Haar scaling function, to obtain smoother scaling functions.

# 6 The $su_q(2)$ limit of $\mathcal{A}_{j_0,j}$

The quantum group $su_q(2)$, or its extensions [21-26,28-31], are associative algebras over $C$ generated by three operators, $J_0 = (J_0)^\dagger$, $J_+$, and $J_- = (J_+)^\dagger$, satisfying the commutation relations

$$[J_0, J_+] = J_+, \qquad [J_0, J_-] = -J_-, \qquad [J_+, J_-] = [J_0]_s, \qquad (52)$$

where $[J_0]$ is a deformation of $J_0$, that is, a real parameter-dependent function of $J_0$, holomorphic in the neighbourhood of zero, and approaching $2J_0$ in the limit $s \to 0$. We have to find a deforming functional that transforms the $\mathcal{A}_{j_0,j}$ generators into operators satisfying the commutation relations eq.(52). For this the first equation in eqs.(33) can be written as

$$(j_0 - \mathcal{G}(s, J_0))j_+ = j_+ j_0, \tag{53}$$

and hence, for every entire function $p(\xi)$, we can write

$$p(j_0 - \mathcal{G}(s, j_0))j_+ = j_+ p(j_0). \tag{54}$$

Let us consider the functional equation

$$\Gamma(\xi - \mathcal{G}(s, \xi)) = \Gamma(\xi) - 1 \tag{55}$$

for a given function $\Gamma(\xi)$ with $\xi$ generic. If this equation has a solution $\Gamma(\xi)$ that is an entire function, then eq.(55) can be written, for $p(\xi) = \Gamma(\xi)$, in the form $[\Gamma(j_0), j_+] = j_+$, and correspondingly $[\Gamma(j_0), j_-] = -j_-$. These equations give the correspondence between the wavelet algebras and the q-deformed algebras like $su_q(2)$ or other extensions of them. They map the structure of $\mathcal{A}_{j_0,j_\pm}$, eq.(33), onto the structure described by the first two equations in eq.(52). For the third commutator of each of these algebras we use the function $\Gamma(\xi)$, which allows eqs.(33) to be reduced to the third equation in eqs.(52) through the mapping:

$$J_0 = \Gamma(j_0), \qquad J_\pm = j_\pm, \tag{56}$$

If $\Gamma$ is invertible, the third equation in eqs.(33) in $\mathcal{A}_{j_0,j}$ can be reduced to the third equation of eqs.(52) of an algebra defined by the function $[J_0]$ with the identification $\mathcal{F} \circ \Gamma^{-1} = [\ ]$ where $[\ ] \circ \Gamma$ means the composition of the two functions, i.e., $([\ ] \circ \Gamma)(\xi) = [\Gamma(\xi)]$. In the case of $su_q(2)$, i.e. $[J_0[= [2J_0]_q$, the function $\mathcal{F}(s, j_0, s)$ becomes:

$$\mathcal{F}(s, j_0) = -\frac{(\phi(j_0))^2 - (\phi(j_0))^{-2}}{q - q^{-1}}, \tag{57}$$

where $q = e^s$, $\phi(\xi) \equiv q^{-\Gamma(\xi)}$ has to satisfy the equation $\phi(\xi - \mathcal{G}(s, \xi)) = q\phi(\xi)$. Consequently, eqs.(53-55) provide the reduction of $su_q(2)$ into $\mathcal{A}_{j_0,j}$ through $\Gamma$. For details of the above technique of mapping and reduction of non-linear algebras one can see examples in [31]. In the case of the algebra $\mathcal{A}_{1,1}$, eqs.(53-55) give the condition

$$p(W_0 - 1) = W_0 - G(1, W_0), \tag{58}$$

which has the solution

$$J_0 = p(W_0)\Big|_{\mathcal{A}_{1,1}} = cosh(2^{-W_0}b), \tag{59}$$

for any arbitrary constant $b$. By inverting eq.(59) we obtain

$$W_0 = \mp \frac{1}{ln(2)} ln(\frac{\partial}{b}). \tag{60}$$

Conversely, starting from $su_q(2)$ and its deformation $[\ ]_q$ one can map it into $\mathcal{A}_{1,1}$ where for the operator $F(s,T) = F(s, T(W_0))$ we have

$$F = f(\Gamma(W_0)) = f(p^{-1}(W_0)) = [2p^{-1}(W_0)]. \tag{61}$$

A diagram containing these algebraic morphisms is presented in Fig.3.

# 7   Further extensions, comments and conclusions

In this paper we present a method for connecting classes of non-linear (q-deformed) algebras with scaling functions and wavelets. The connection is reciprocal. The algebra $\mathcal{A}_{j_0,j}|_{s=1}$ allows the choice of different dilation equations, i.e., for different functions $j(1,T)$ one can obtain different scaling functions. The function $j_0(1,T)$ should be choosen in order to give the correct wavelet function through the action of $j_0 j_-$. The connection with quantum groups is more transparent if we choose for $j_0 = [\partial]_{s=1/2} = W_0(2, 1/2)$, eq.(14), and for $\Phi$, the Haar scaling function, $j_- = D(1 + T^{-1})$. The action of the q-derivative $[\partial]_{1/2}$ on $\Phi$ gives exactly the Haar wavelet. In this example the operator $j_0$ is anti-Hermitian and we can identify its spectrum (imaginary eigenvalues) from the commutation relations. We also have the relation $\Psi = (j_0 - \tilde{G}(1, j_0))\Phi$, so that one can express the eigenvectors $|a>$ of $J_0$ in the wavelet basis of the algebra. If $|a> = \sum_{j,n} C_{j,n}\Phi_{j,n}$, we obtain a recursion relation for the coefficients of $|a>$ and for the eigenfunctions of $j_0$ $\sum_k C_{j,p-2^j k}\mathcal{C}_k = aC_{j,p}$ for any $j$ and $p$ integer, where $\mathcal{C}_k$ are the Taylor coefficients of the function $-\xi^{-\lambda}g(-\xi)$. We note that by using the q-derivative instead of a function of T we obtain exactly the same mother wavelet (Haar wavelet).

We note a natural extension of the above developments coming from a realization of $sp(2, R)$ in terms of the full algebra of the symmetry of the real line: $X_0 = x\partial$ (dilations), $X_- = \partial$ (translations), and $X_+ = x^2\partial$ (expansions). This is the maximal finitely generated real Lie algebra with generators in the form $x^n\partial$. By exponentiation the generators $X_0, X_-$ provide $D$ and $T$ and the last generator has the action $E^a = e^{aX_+}f(x) = f(x/1 - ax)$. For $a > 0$ the action corresponds to a contraction of the function and for $-1 < a < 0$ it is a dilation. But for $a < -1$ and $|a| \geq 1$ we obtain a splitting of the function into two Heaviside distributions: $E^a f(x) = H(x) + H(-x + 1/a)$. A possible closed non-linear algebra containing $D, T$ and $E$ might be an opportunity for the application of wavelets with multi-scale properties [18,19].

In conclusion, we have found that the operators of dilation and translation $(D,T)$ can be combined in such a way as to generate non-linear algebras which depend on certain parameters $(s,\alpha)$. We have investigated these algebras from the point of view of quantum groups, discussing their unirreps, Casimir operators, and reductions of these algebras to other q-deformed algebras, like $su_q(2)$ or to the Fourier series generating algebra. It has been shown that such algebras provide an appropriate framework for the foundation of wavelet analyses and for the obtaining the corresponding scaling functions. We have worked out two examples: the Haar and the B-scaling functions. The algorithm provides a general algebraic method for finding a specific scaling function. Direct applications of such an approach can be found in the theory of finite-difference equations and q-difference equations [21-25,27].

This study represents only a first step towards understanding the relation between the theory of non-linear algebras (having exponential spectra) and wavelets with their finite-difference equations. We conjecture that such an approach to scale invariant structures can lead to interesting mathematical constructions and tools for identifying isolated coherent structures associated with certain mother wavelets and limiting non-linear algebraic struc-tures. The transition between such self-organized structures can be carried out through the modification of the deformation parameter $q$, with the intermediate domain representing "noise" (non-closed algebraic structures).

FIGURE CAPTIONS

1. The dependence on $a_0$ of three eigenvalues $a_n$ of $W_0$ in $\mathcal{A}_{1,1}$, for $n = 2$ (the largest period), 4 and 6 (the smallest period). The three spectra have self-similar structure for $a_0 \in [0, 1]$.

2. The continuous deformation of the scaling function $\Phi$ in $\mathcal{A}_{s,\alpha}$ as a function of the parameter $s$. For $s = 0, 1$ this is the Haar scaling function.

3. A diagrammatic representation of the algebraic maps (arrows) and morphisms ($\simeq$) of $\mathcal{A}_{j_0 j}$, $\mathcal{A}_{s,\alpha}$, $su_q(2)$, $su(2)$ and $\mathcal{F}_{0,1}$.

# References

[1] I. Daubechies, *Commun. Pure. Appl. Math* **49**, 909 (1988); I. Daubechies and A. Grossmann, *J. Math. Phys.* **21**, 2080 (1980);

[2] I. Daubechies, *Ten Lectures on Wavelets* (SIAM, Philadelphia, 1992); C. K. Chui, *An Introduction to Wavelets* (Academic Press, New York, 1992); G. Kaiser, *A Friendly Guide to Wavelets* (Birkhäuser, Boston, 1994); Y. Meyer in *Wavelets*, ed. J. M. Combes, A. Grossmann and Ph. Tchamitchian (Springer, Berlin, 1988), p1.

[3] Ph. Tchamitchian and B. Torresani in *Wavelet and their applications*, ed. M. B. Ruskai (Jones amd Bartlett Publ. Boston, 1992); G. Kaiser, *SIAM J. Math. Anal.* **23**, 222 (1992); C. K. Chui and A. Cohen in *Approximation Theory VII*, eds. E. W. Cheuey, C. K. Chui and L. L. Schumaker (Academic Press, Boston, 1993)

[4] J. J. Benedetto, M. W. Frazier, *Wavelets: Mathematics and Applications* (CRC Press, Boca Raton, 1994); H. Ogawa and N-E. Berrached in *Contemporary Mathematics* **190**, eds. M. E. H. Ismail (Am. Math. Soc., Providence, Rhode Island, 1995); G. Strang, *Physics D* **60**, 239 (1992); *SIAM Rev.* **31**, 614 (1989).

[5] D. Han, Y. S. Kim and M. E. Noz, *Phys. Lett. A* **206**, 299 (1996); Y. S. Kim and E. P. Wigner, *Phys. Rev. A* **36**, 1293 (1987);

[6] C. Meneveau, *Phys. Rev. Lett.* **66**, 1450 (1991); M. Ima and S. Toh, *Phys. Rev. E* **52**, 6189 (1995); A. Grossman and J. Morlet, *SIAM J. Math. Anal.* **15**, 723 (1984); R. Everson, L. Sirovich and K. R. Sreenivasan, *Phys. Lett. A* **145**, 314 (1990); R. Benzi and M. Vergassola, *Fluid Dyn.* **8**, 117 (1991); A. Scotti and C. Meneveaux, *Phys. Rev. Lett* **78**, 867 (1997)

[7] C. Meneveau and K. R. Sreenivasan, *J. Fluid. Mech* **224**, 429 (1991); J. F. Muzy, E. Bacry and A. Arneodo, *Phys. Rev. Lett.*, **67** (1991) 3515; M. Greiner, J. Giessmann, P. Lipa and P. Carruthers, *Z. Phys. C* **69**, 305 (1996)

[8] A. Krolak and P. Trzaskoma, *Class. and Quant. Grav.* **13**, 813 (1996); J. Widjaja, Y. Tomita and A. Wahab, *Opt. Comm.* **132**, 217 (1996); J. Pando and L. Z. Fang, *Astrophys. J.* **459**, (1996).

[9] Z. Huang, I. Sarcevic, R. Thews and X. -N. Wang, *Phys. Rev. D* **54** 750 (1996).

[10] J. Kantelhardt, M. Greiner and E. Roman, *Physica A* **220**, 219 (1995); H. E. Roman, J. W. Kantelhardt and M. Greiner, *Europhys. Lett.* **35**, 641 (1996)

[11] K. Cho, T. A. Arios and J. D. Joannopoulos, *Phys. Rev. Lett.* **71**, 1808 (1993)

[12] F. Bagarello, *J. Phys. A: Math. Gen.* **29**, 565 (1996)

[13] M. Farge and G. Rabreau, *C. R. Acad. Sci. Paris, Ser. II* **307**, 1479 (1988)

[14] W. Tobocman, Inv. Probl. **12**, 499 (1996)

[15] S. Afanasiev, M. Altaisky and Yu. Zhestkov, *Il Nuovo Cimm.* **108**, 919 (1995)

[16] J. P. Modisette, P. Nordlander, J. L. Kinsey and B. R. Johnson, *Chem. Phys. Lett.* **250**, 485 (1996); D. Mugnai, A. Ranfagni, R. Ruggeri and A. Agresti, *Phys. Rev. E* **50**, 790 (1994); G. Berkovz, J. Elezgaray and P. Holmes, *Physica D*, **61**, 47 (1992); A. O. Barut, *Found. Phys.* **20**, 1233 (1990)

[17] I. Makoto and T. Sadayoshi, *Phys. Rev. Lett.* **52**, 6189, (1995); L. Gagnon and J. M. Lina, *J. Phys. A: Math. Gen.* **27**, 8207 (1994)

[18] A. Ludu and J. P. Draayer in *multi-scale Phenomena in Physics and Engineering*, Eds. P. Vashishta and R. Kalia, (Louisiana State University, Febr. 7-9, 1997); A. Ludu and J. P. Draayer, *Hamiltonian System and Symmetry Algebra for Wavelets*, submited; A. Ludu and J. P. Draayer, Int. Conf. *Physics at The Turn of Millenium*, 22-26 Oct. 1996, Santa Fe, NM, (World Scientific, Singapore,in print).

[19] A. Ludu and M. Greiner, *ICTP - preprint* **IC/95/288**.

[20] N. I. Vilenkin, *Special Functions and The Theory of Group Representations*, (Nauka, Moscow, 1955); J. D. Talman, *Special Functions. A Group Theoretic Approach*, (W. A. Benjamin, Inc., New York, 1968).

[21] V. Chari and A. Pressley *A Guide to Quantum Groups* (Cambridge U.P., Cambridge, 1994); J. Fröchlich and T. Kerler, *Quantum Groups, Quantum Categories and Quantum Field Theory* (Springer, Berlin, 1993); E. Witten, *Nucl. Phys. B* **330**, 285 (1990)

[22] D. Levi, L. Vinet and P. Winter, eds., *Symmetries and Integrability of Difference Equations* (American Mathematical Society, Providence, 1997).

[23] A. Ludu and W. Greiner, *Found. Phys.*, to appear.

[24] R. Floreanini and L. Vinet, *J. Math. Phys.* **36**, 7024 (1995)

[25] R. Floreanini and L. Vinet, *J. Math. Phys.* **36**, 3134 (1995).

[26] A. J. Macfarlane, *J. Math. Phys. A: Math. Gen.* **22**, 4581 (1989).

[27] C. R. Adams, *Bull. Am. Math. Soc.* **37**, 361 (1931); G. E. Andrews, *q-Series: Their development and applications in analysis, number theory, combinatorics, physics, and computer algebra* (Am. Math. Soc., Providence, Rhode Island, 1986)

[28] A. P. Polychronakos, *Mod. Phys. Lett. A* **5**, 2325 (1990).

[29] M. Rocek, *Phys. Lett. B* **255**, 554 (1991).

[30] A. Ludu and R. K. Gupta, *J. Math. Phys.* **34**, 5367 (1993).

[31] C. Delbeque and C. Quesne, *J. Phys. A: Math. Gen.* **26**, L127 (1993); D. Bonatsos, C. Daskaloyannis, P. Kolokotronis, A. Ludu and C. Quesne, *J. Math. Phys.* **38**, 369 (1997).

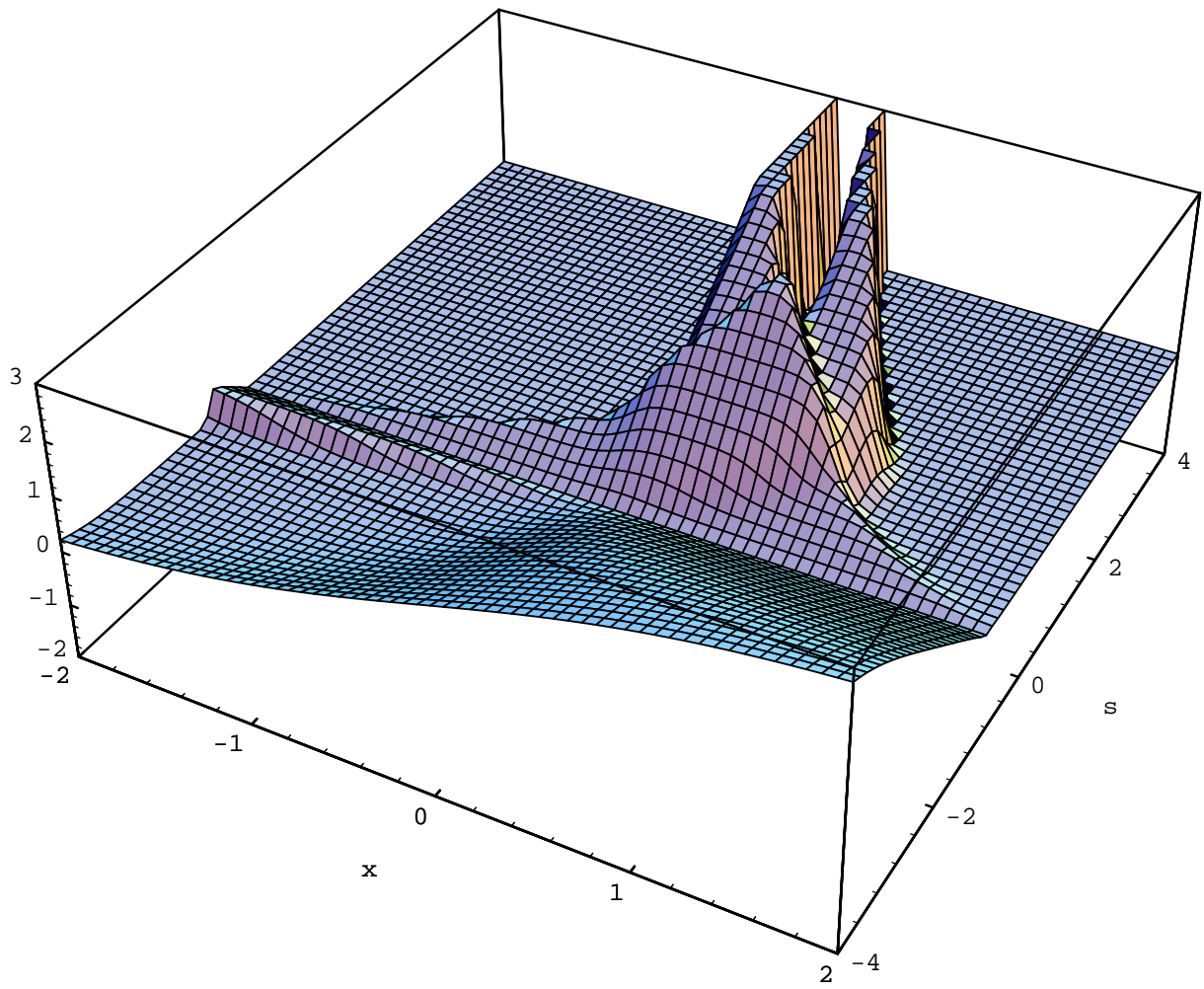

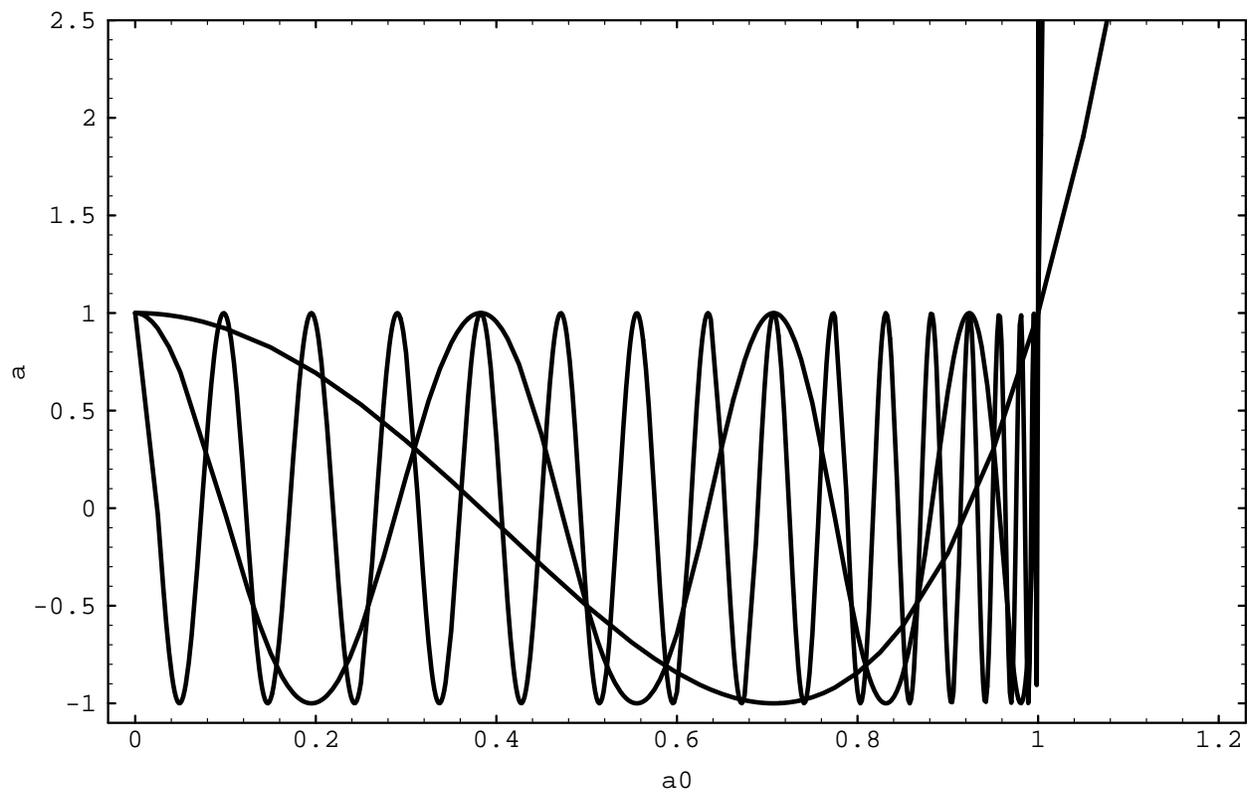

$$
\begin{array}{ccccc}
 & & \mathcal{A}_{2,1/2} & \xrightarrow{T,D} & \Psi_H \\
 & & \uparrow & \swarrow \mathcal{F}(W_0) \;\; \Big\uparrow {\scriptstyle -\partial+i\pi} & \\
\mathcal{A}_{j_0,j} & \longrightarrow & \mathcal{A}_{s,\alpha} & \longrightarrow & \mathcal{A}_{1,1}(\Phi_H) \\
\Big\downarrow {\scriptstyle p} & \nearrow {\scriptstyle p} & & \searrow & \\
sl_q(2) & \xrightarrow{q=1} & sl(2) & \mathcal{A}_{0,2} & \\
\Big\downarrow {\scriptstyle q=-1} & \searrow {\scriptstyle q=i} & & \swarrow & \\
sp(2,R) \simeq \mathcal{F}_0 & & e(2) \simeq \mathcal{F}_1 & &
\end{array}
$$